\documentclass[prb,twocolumn,showpacs,preprintnumbers,amsmath,amssymb]{revtex4-1}

\usepackage{graphicx}
\usepackage{dcolumn}
\usepackage{bm}

\begin{document}

\preprint{APS/123-QED}

\title{Mach number dependence of electron heating \\
in high Mach number quasiperpendicular shocks}

\author{Shuichi Matsukiyo}
 \email{matsukiy@esst.kyushu-u.ac.jp}
\affiliation{%
Department of Earth System Science and Technology, Kyushu University\\
6-1 Kasuga-Koen, Kasuga, 816-8580, Fukuoka, Japan
}%


\date{\today}

\begin{abstract}
Efficiency of electron heating through microinstabilities 
generated in the transition region of a 
quasi-perpendicular shock for wide range of Mach numbers 
is investigated by utilizing PIC 
(Particle-In-Cell) simulation and model analyses. In the 
model analyses saturation levels of effective electron 
temperature as a result of microinstabilities 
are estimated from an extended quasilinear (trapping) analysis 
for relatively low (high) Mach number shocks. Here, MTSI 
(modified two-stream instability) is assumed to become 
dominant in low Mach number regime, while BI (Buneman 
instability) to become dominant in high Mach number regime, 
respectively. It is revealed that Mach number dependence of 
the effective electron temperature in the MTSI dominant case 
is essentially different from that in the BI dominant case. 
The effective electron temperature through the 
MTSI does not depend much on the Mach number, although that 
through the BI increases with the Mach number as in the 
past studies. The results 
are confirmed to be consistent with the PIC simulations 
both in qualitative and quantitative levels. The model 
analyses predict that a critical 
Mach number above which steep rise of electron heating rate 
occurs may arise at the Mach number of a few tens. 
\end{abstract}

\pacs{52.50.Lp, 52.35.-g}
\keywords{electron heating, collisionless shock, microinstability}
\maketitle

\section{\label{intro}Introduction}

Efficiency of electron heating in high Mach number collisionless 
shocks is one of the outstanding issues of space plasma physics 
as well as astrophysics. It is known from observations that 
quite efficient heating occurs in extremely high Mach number 
shocks like SNR (super nova remnant) shocks \cite{par06,bad05,bad03,hug00}, 
while in-situ observations of earth's bow shock mostly show 
inefficient heating of electrons \cite{mos85}. 
This is probably due to the fact that a dominant electron heating 
process and its efficiency through a variety of microinstabilities 
in shock transition regions strongly depend on Mach numbers. 
Typical Mach number of the earth's bow shock is $< 10$, 
while that of the SNR shocks is of the order of $100 - 1000$. 
In the late 1980s a new process of strong and rapid electron heating 
in extremely high Mach number shocks was proposed \cite{pap88,car88}. 
Refs. 6 and 7 discussed the so-called two step 
instabilities generated by a high velocity reflected ion 
beam which destabilizes BI (Buneman instability) followed by 
ion acoustic instability. A necessary condition for the BI to 
be destabilized in the foot of quasi-perpendicular shocks is 
$M_A / \sqrt{\beta_e} > 20 \sim 30$ \cite{car88,mat03}, 
where $M_A$ is the Alfv\'{e}n Mach number and 
$\beta_e = 8 \pi n T_e / B^2$ the electron plasma beta. 
If the above condition is satisfied, rapid electron heating 
with saturation temperature $\propto M^2_A$ occurs and the ion 
acoustic instability sets in leading to 
further electron heating \cite{pap88, car88}. This 
condition is easily satisfied in SNR shocks, although it is 
usually not in the earth's bow shock. 

The two step instabilities were again shed light on after 
Ref. 9 by utilizing electromagnetic 
full PIC (particle-in-cell) simulations. They improved 
understandings of the process not only in terms of electron 
heating but also in terms of a production process of 
non-thermal electrons, and promoted a number of subsequent 
simulation studies \cite{hos02,shi04,die04,shi05,mcc05,die06,
ama07,ohi07,ohi08,ume08,die09,ama09a,ama09b}. As a result, 
understandings of electron heating or acceleration processes 
initiated by the BI in the context of the shock physics 
have been extensively developed.

On the other hand, it has been well-known that various 
microinstabilities are possible to get excited also in 
relatively low Mach number supercritical shocks 
\cite{wu84,pap85}. 
Since most of the microinstabilities are inseparable with 
electron dynamics, their nonlinear evolutions in self-consistently 
reproduced shock structures in PIC simulations are studied 
only recently \citep{sch03,sch04,mus06,mat06a} except 
for Refs. 29-31. 
It is revealed that MTSI (modified two-stream instability) 
becomes dominant in the foot of the earth's bow shock or 
interplanetary shocks at $\sim1$ AU \citep{sch03,sch04,mat06a}. 
Simulation studies on nonlinear evolutions of the MTSI in 
periodic systems were extensively studied in the 1970s and 1980s 
\cite{mcb72,ott72,tan83}, although these simulations 
imposed some strong assumptions that the code was 
electrostatic \cite{mcb72,ott72} and unmagnetized ions with 
small ion-to-electron mass ratio were used 
\cite{mcb72,ott72,tan83}. Recently, some developement without 
these assumptions are made \cite{mat03,mat06b}. 
Ref. 8 showed in a one-dimensional 
simulation with realistic mass ratio that long time evolution 
of the MTSI results in lower cascade 
of wave spectrum and associated electron heating. Furthermore, 
it is indicated in two-dimensional simulation that the MTSI 
can finally survive even if other possible instability like 
electron cyclotron drift instability is present \cite{mat06b}. 
All these simulation studies show electron heating or 
acceleration in the long time evolution of the system. 
However, as mentioned already, electron heating in the 
earth's bow shock is seldom observed in-situ.

Such a discripancy may arise because of lack of systematic 
estimate of electron heating rate in simulation studies with 
more realistic parameters. For example, all of the above 
simulation studies on the MTSI assume very low values of 
squared ratio of electron plasma to cyclotron frequencies, 
$\tau = \omega^2_{pe} / \Omega^2_e\leq 10$. Importance of 
$\tau$ in association with electron heating in the long time 
evolution of the MTSI is unresolved, while $\tau$ is known 
to be a crucial parameter controlling nonlinear electron 
heating and acceleration processes in the BI dominant system 
\cite{shi04}. Electron plasma beta is also rather small in 
most of the previous studies ($\beta_e \leq 0.1$). This may 
exaggerate effects of electron trapping in the nonlinear 
stage of the MTSI \cite{mcb72,sch03,mat03,sch04,mat06a}. 
The electron plasma beta in the solar wind 
is usually several times higher so that importance of the trapping 
process may be blurred. Hence, in realistic situations expected 
to be achieved in the typical solar wind conditions an alternative 
electron heating mechanism and its efficiency are ought to 
be considered. 

In this paper, first, electron heating 
through the MTSI in transition regions of high Mach 
number quasi-perpendicular shocks is reconsidered in 
section II. 
Here, quasilinear diffusion is assumed 
as an alternative electron heating process. 
The analysis assumes a broad wave spectrum and small 
wave amplitudes which are likely to be satisfied in 
the past simulation studies mentioned above 
\cite{mcb72,mat03}. An additional assumption that 
relative phases among different wave modes are random 
is also imposed, while this point is currently 
not evident. Under these assumptions, 
taking second order velocity moments of the so-called 
quasilinear equations provides evolution equations of 
electron kinetic energies. These equations are numerically 
integrated by imposing the quasilinear assumptions in 
every instantaneous time steps to obtain saturation 
kinetic energies as a fuction of the Mach number of shocks. 
This approach allows to discuss long time evolution of 
macroscopic quantities including the electron temperature 
without any restrictions for parameters. It is intriguing 
whether the Mach number dependence of electron heating throug 
the MTSI in relatively low Mach number regime is different 
from that through the BI in extremely high Mach number 
regime. If that is the case, one may expect presence of a 
critical Mach number above which electron heating rate runs 
up due to the switching of the dominant microinstability 
\cite{hos02}. In section III, the 
results are compared with 1D PIC simulations in which the 
MTSI is dominantly generated in the foot and 
effective electron temperature just behind of the 
shock is measured. Furthermore, post-shock effective 
electron temperature is measured also in cases 
that the BI gets excited in the foot. In those 
cases a rough estimate of the electron temperature is 
given by using the well-known trapping theory. 
Finally, a summary and discussions including 
estimate of the critical Mach number are given 
in section IV.

\section{\label{eqlana}Extended Quasilinear Analysis on Electron 
Heating through MTSI}

\subsection{System Configurations}

A foot region of a quasiperpendicular shock is 
modeled by a plasma composed of incoming electrons 
at rest, incoming ions, and specularly reflected 
ions with local approximation. The ambient magnetic 
field is along the $z$-axis, while the incoming 
and the reflected ion beams are streaming parallel 
and antiparallel to a wavevector which is defined 
in the $x-z$ plane and parallel to the shock normal. 
This is essentially the same system discussed in 
the linear analysis of Ref. 8. Since the 
MTSI grows much faster than ion cyclotron period, ions 
are assumed to be unmagnetized. 

\subsection{Governing Equations}

Saturation levels of electron kinetic energy 
through the MTSI are examined as follows. In a Vlasov equation,
\begin{equation}
\label{vlasov}
 {\partial f_j \over \partial t} + {\bf v} \cdot \nabla f_j 
 + {q_j \over m_j} \left( {\bf E} + { {\bf v \times B} 
 \over c} \right) \cdot \nabla_{\bf v} f_j = 0,
\end{equation}
variables are expanded around zero-th order slowly 
varying quantities (with subscript 0) as 
\begin{equation}
\label{field}
\begin{array}{ccccc}
 f_j & = & F_{j0} & + & f_{j1}, \\
 {\bf B} & = & {\bf B_0} & + & {\bf B_1}, \\
 {\bf E} & = & & & {\bf E_1},
\end{array}
\end{equation}
where $f_j$, $q_j$, and $m_j$ are a distribution function, 
electric charge, and mass of species $j$, $c$ is speed of 
light, 
${\bf B}$ and ${\bf E}$ represent magnetic and electric 
fields, respectively. The variables with subscript 1 
denote fluctuations corresponding to linear oscillations. 
With a standard approach of quasilinear analysis, 
an evolution equation of $F_{j0}$ is obtained 
as follows \cite{sti92,ken66}. 
\begin{align}
\label{qleq}
 {\partial F_{0} \over \partial t} = 
 \lim_{V \rightarrow \infty} &{\pi q^2 \over m^2} 
 \sum^{\infty}_{n=-\infty} \int {d^3 {\bf k} \over V} \notag \\
 &L v_{\perp} \delta(\omega_{r,{\bf k}} - k_{\parallel} 
 v_{\parallel} - n \Omega) |\Psi_{n,{\bf k}}|^2 v_{\perp} 
 L F_0
\end{align}
Here, $V$ is the volume of integration, 
$\Omega = q B_0 / m c$ the cyclotron frequency, 
$v_{\parallel}$ and $v_{\perp}$ the velocity 
components parallel and perpendicular to ${\bf B_0}$, 
${\bf k}$ the wavevector which is also composed 
of parallel and perpendicular components 
$(k_{\parallel}, k_{\perp})$, and $\omega_{r,{\bf k}}$ 
indicates 
real wave frequency which is a function of ${\bf k}$ 
and satisfies the linear dispersion relation. 
The subscript $j$ has been eliminated for simplicity. 
Only resonant wave-particle interactions corresponding 
to zero arguments of the delta function, 
$\omega_{r,{\bf k}} - 
k_{\parallel}  v_{\parallel} - n \Omega = 0 \: (n = 0, 
\pm 1, \pm 2, \cdots)$, are taken 
into account. Moreover, 
\begin{equation}
 L = {1 \over v_{\perp}} \left[ 
 \left( 1 - {k_{\parallel} v_{\parallel} \over
 \omega_r} \right) {\partial 
 \over \partial v_{\perp}} + {k_{\parallel} v_{\perp} 
 \over \omega_r} {\partial 
 \over \partial v_{\parallel}} \right],
\end{equation}
%
\begin{align}
\label{psi2}
|\Psi_{n,{\bf k}}|^2 = &{E^2_{1,{\bf k}} \over 1 + A^2_{\perp} + 
A^2_{\parallel}} \left[ {(1+A_{\perp})^2 \over 4} 
J^2_{n-1}(\zeta) \right. \notag \\
&\left. + {(1-A_{\perp})^2 \over 4} J^2_{n+1}(\zeta) 
+ {v^2_{\parallel} \over v^2_{\perp}} A^2_{\parallel} 
J^2_n(\zeta) \right],
\end{align}
where $A_{\parallel} = E_{1z,{\bf k}} / E_{1x,{\bf k}}, 
A_{\perp} = i E_{1y,{\bf k}} / E_{1x,{\bf k}}$, 
$E^2_{1,{\bf k}} = E^2_{1x,{\bf k}} + E^2_{1y,{\bf k}} + 
E^2_{1z,{\bf k}}$, and $J_n(\zeta)$ represents the $n-$th 
order Bessel function with argument $\zeta = k_{\perp} 
v_{\perp} / \Omega$. When the above expression of 
$\Psi_{n,{\bf k}}$ is derived, the linear dispersion relation 
is taken into account. Note that in 
eqs.(\ref{qleq})-(\ref{psi2}) 
parallel and perpendicular directions 
are defined in terms of ${\bf B_0}$. Although this 
is true for electrons, further note is needed for 
ions. Because ions are assumed to be unmagnetized, 
distinction of parallel and perpendicular directions 
in this sense is meaningless. However, their evolution 
equations can be formally written in a similar form 
by defining parallel and perpendicular directions 
in terms of the ion beam velocity. This corresponds to a 
transformation for the rotation of the susceptibility 
tensor \cite{mat03}.

To discuss the extended quasilinear evolution of kinetic 
energies through the MTSI, we examine parallel and 
perpendicular kinetic energy densities defined by 
\begin{equation}
\label{kpara}
 K_{\parallel}(t) = {m \over 2} \int d^3 {\bf v} 
 (v_{\parallel} - v_0)^2 F_0,
\end{equation}
%
\begin{equation}
\label{kperp}
 K_{\perp}(t) = {m \over 2} \int d^3 {\bf v} 
 v^2_{\perp} F_0.
\end{equation}
After some algebra, the following evolution equations 
of the kinetic energies are obtained.
\begin{align}
\label{paraene}
 {\partial K_{\parallel} \over \partial t} &= 
 \lim_{V \rightarrow \infty} {2 \pi^2 q^2 \over m} 
 \sum_n \int {d^3 {\bf k} \over V} \left[ 
 {\nu \over 
 \omega_{r,{\bf k}} |k_{\parallel}|} \right. \notag \\
& \left. \left( 
 {n \Omega \over \omega_{r,{\bf k}} v^2_{t\perp}} 
 + {\nu \over 
 \omega_{r,{\bf k}} v^2_{t\parallel}} \right) 
 \int dv_{\perp} v^3_{\perp} (|\Psi_n|^2 F_0 
 )_{\delta=0} \right]
\end{align}
%
\begin{align}
\label{perpene}
 {\partial K_{\perp} \over \partial t} &= 
 \lim_{V \rightarrow \infty} {2 \pi^2 q^2 \over m} 
 \sum_n \int {d^3 {\bf k} \over V} \left[ 
 {n \Omega \over \omega_{r,{\bf k}} |k_{\parallel}|} \right. \notag \\
& \left. \left( {n \Omega \over \omega_{r,{\bf k}} v^2_{t\perp}} 
 + {\nu \over 
 \omega_{r,{\bf k}} v^2_{t\parallel}} \right) 
 \int dv_{\perp} v^3_{\perp} (|\Psi_n|^2 F_0 
 )_{\delta=0} \right]
\end{align}
Here, $\nu \equiv \omega_{r,{\bf k}} - k_{\parallel} 
v_0 -n \Omega$, and $()_{\delta=0}$ means a corresponding 
value at which the argument of the delta function is 
zero. The spectral energy density of the electric 
field fluctuations evolves as 
\begin{equation}
\label{waveene}
 {\partial E^2_{\bf k} \over \partial t} = 2 
 \gamma_{\bf k} E^2_{\bf k},
\end{equation}
where $\gamma_{\bf k}(t)$ denotes the linear growth rate 
of the wave mode having wavevector ${\bf k}$. 
Conservation of total energy is written as 
\begin{equation}
\label{enecon}
 \sum_j \left( {n_j m_j \over 2} v^2_{j0} + K_{j\parallel} 
 + K_{j\perp} \right)
 + \sum_{\bf k} {E^2_{1{\bf k}} \over 8 \pi} = const.
\end{equation}
Here, $n_j$ and $v_{j0}(t)$ denote density and bulk velocity 
of species $j$. Note that $\partial K_{\perp} / \partial t = 0$ 
for unmagnetized ions and $v_{e0} = 0$. Further, 
$\omega_{r,{\bf k}}$ and $\gamma_{\bf k}$ are defined at 
an instantaneous time. 
Eqs.(\ref{paraene})-(\ref{enecon}) form 
a closed set of extended quasilinear evolution equations 
of the system. We solve the above set of equations 
numerically in the following subsection.

Similar analyses were performed by many authors 
\cite{mcb72,dav75,ish83a,ish83b}. 
Ref. 38 introduced detailed 
procedure of the method and applied it to estimate 
of saturation of electromagnetic ion cyclotron 
instability driven by ion temperature anisotropy. 
Refs. 39 and 40 
directly solved the quasilinear 
equation to discuss electron heating and formation 
of a high energy ion tail through the ion acoustic 
instability. Ref. 33, 
on the other hand, discussed a 
problem similar to here on the MTSI and gave detailed 
comparisons of their results 
with 1d and 2d PIC simulations. However, their analysis 
includes only nonresonant wave-particle interactions which 
in turn only slight electron heating which is much less 
effective compared with a subsequent heating process 
through nonlinear trapping observed in their PIC 
simulation. Furthermore, they 
imposed an electrostatic assumption on whole 
their analyses as well as on their PIC 
simulations. As they pointed out, the electrostatic 
approximation is valid only when $v_{j0} / v_A < (1 + 
\beta_e )^{1/2}$, where $v_A$ denotes 
the Alfv\'{e}n velocity, and $\beta_e$ is the electron 
plasma beta (ratio of electron thermal to magnetic pressures). 
Usually in solar wind $\beta_e$ 
is of the order of $0.1-1$ so that the above 
condition reads $v_{j0} / v_A \stackrel{<}{\sim} 1$ 
which is probably satisfied only in transition 
regions of rather low Mach number shocks. Since 
our attention is paid for wide range of Mach 
numbers, we exclude such an assumption. 

Before solving eqs.(\ref{paraene})-(\ref{enecon}) 
numerically, a few assumptions are imposed. 
The distribution functions of each species remain 
in (shifted) Maxwellian all the time after 
Ref. 38. By this 
assumption, it is easy to obtain $\omega_{r,{\bf k}}(t)$ 
and $\gamma_{\bf k}(t)$ at each time step as 
instantaneous solutions of 
the linear dispersion relation. 
The actual distribution functions are given by
\begin{equation}
 F_{j0} = {n_{j} \over (2 \pi)^{3/2} v_{tj\parallel} 
 v^2_{tj\perp}} \exp \left(- {(v_{j\parallel} - 
 v_{j0})^2 \over 2 v^2_{tj\parallel}} - {v^2_{j\perp} 
 \over 2 v^2_{tj\perp}} \right)
\end{equation}
for species $j$. 
Here $v_{tj\parallel}=\sqrt{T_{j\parallel} / m_j}$ and 
$v_{tj\perp}=\sqrt{T_{j\perp} / m_j}$ denote parallel and 
perpendicular thermal velocities. As mentioned before, 
note that the parallel and perpendicular directions 
are difined in association with ${\bf B_0}$ for 
electrons, while with ${\bf v_{j0}}$ 
for incoming and reflected ions. 
The bulk velocities of the incoming and reflected ions 
(subscripts $j=i$ and $r$, respectively) satisfy zero 
current condition so that $n_i v_{i0} + n_r v_{r0} = 0$. 
It is also assumed that 
the response of the ions to the wave fields is 
electrostatic, while the wave-electron interactions 
are treated as fully electromagnetic. Therefore, the 
ion heating is thought to be parallel to ${\bf v_{i0}}$ 
or ${\bf v_{b0}}$.

Under these assumptions, we obtain the 
following evolution equations of kinetic energies. 
\begin{align}
\label{eneea}
 {\partial K_{e\parallel} \over \partial t} &= 
 \lim_{V \rightarrow \infty} {2 \pi^2 e^2 \over m_e} 
 \sum_n \int {d^3 {\bf k} \over V} \left[ 
 {\nu_e \over 
 \omega_{r,{\bf k}} |k_{\parallel}|} \right. \notag \\
& \left. \left( 
 {-n \Omega_e \over \omega_{r,{\bf k}} v^2_{te\perp}} 
 + {\nu_e \over 
 \omega_{r,{\bf k}} v^2_{te\parallel}} \right) 
 \int dv_{\perp} v^3_{\perp} (|\Psi_n|^2 F_{e0} 
 )_{\delta=0} \right]
\end{align}
%
\begin{align}
\label{eneee}
 {\partial K_{e\perp} \over \partial t} &= 
 \lim_{V \rightarrow \infty} {2 \pi^2 e^2 \over m_e} 
 \sum_n \int {d^3 {\bf k} \over V} \left[ 
 {- n \Omega_e \over \omega_{r,{\bf k}} |k_{\parallel}|} 
 \right. \notag \\
& \left. \left( {-n \Omega_e \over \omega_{r,{\bf k}} v^2_{te\perp}} 
 + {\nu_e \over 
 \omega_{r,{\bf k}} v^2_{te\parallel}} \right) 
 \int dv_{\perp} v^3_{\perp} (|\Psi_n|^2 F_{e0} 
 )_{\delta=0} \right]
\end{align}
%
\begin{align}
\label{enei}
 {\partial K_{i\parallel} \over \partial t} = 
 \lim_{V \rightarrow \infty} &{2 \pi^2 e^2 \over m_i} 
 \int {d^3 {\bf k} \over V} \left[ 
 {1 \over |k|} 
 \left( 
 {\nu_i \over 
 \omega_{r,{\bf k}} v_{ti\parallel}} \right)^2 
 \right. \notag \\
& \left. \int dv_{\perp} v^3_{\perp} (|\Psi_0(z=0)|^2 F_{i0} 
 )_{\delta=0} \right]
\end{align}
Here, $\nu_e = \omega_{r,{\bf k}} + n \Omega_e$ and 
$\nu_i = \omega_{r,{\bf k}} - k v_{i0}$, 
respectively. 
In this analysis only the linear resonant 
wave-praticle interactions are taken into account. 
Therefore, change of kinetic energy of the reflected 
ions ($K_{r\parallel}$) can be neglected 
when we consider the MTSI based 
on electron-incoming ion interactions. That is, here, 
the reflected ions are assumed to be a background 
component satisfying current and charge neutralities. 
Reactions of 
the reflected ions may be essential when influences 
of the MTSI on the reformation process are considered 
\cite{sch04}. 
However, it is out 
of scope in this paper. In the 
following we solve eqs.(\ref{eneea})-(\ref{enei}), 
(\ref{waveene}), and 
\begin{equation}
\label{econ}
 {n_i m_i \over 2} v^2_{i0} + K_{i\parallel} 
 + K_{e\parallel} + K_{e\perp}
 + \sum_{\bf k} {E^2_{1{\bf k}} \over 8 \pi} = const.
\end{equation}
instead of eq.(\ref{enecon}). Eq.(\ref{econ}) is 
actually used to determine $v^2_{i0}$. 
Note that contributions 
from the terms $|n|>4$ for $\sum_n$ in eqs.(\ref{eneea}) 
and (\ref{eneee}) are neglected. 
The validity of such an assumption will be checked 
in the following subsection.

\subsection{Numerical Solutions}

\begin{figure}
\includegraphics[width=8cm]{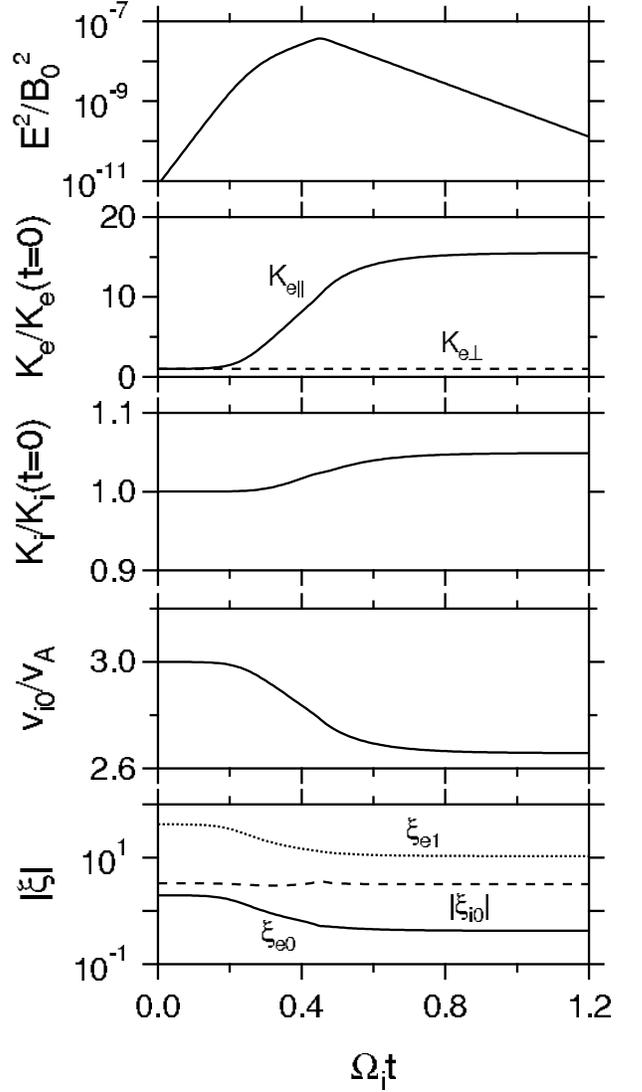}
\caption{Time evolution of the system.
\label{fig01}}
\end{figure}

First, let us overview in Fig.\ref{fig01} time 
evolutions of the variables 
calculated in the system for a typical case. 
The initial parameters are 
$\mu \equiv m_i/m_e = 1836, 
\tau \equiv \omega^2_{pe}/\Omega^2_e = 10^4, 
\beta \equiv 8\pi n_e (T_e + T_i) / B^2_0 \equiv 
\beta_e + \beta_i = 0.4, 
\alpha \equiv n_r/n_i = 1/3, 
\theta_{Bk} (=\Theta_{Bn}) = 85.5^{\circ}$, 
and $M_A \equiv (v_{i0}/v_A) (1+\alpha)/2\alpha = 6$. 
Here, the initial temperatures are isotropic and 
$\beta_e = \beta_i$. $M_A$ and $\alpha$ determine the 
initial $v_{i0}$ 
from the zero current condition. $\theta_{Bk} 
(\Theta_{Bn})$ denotes the angle between the 
ambient magnetic field, ${\bf B_0}$, and the 
wavevector (the shock normal), ${\bf k} ({\bf n})$. The above 
parameters are fixed throughout the analysis if not 
specified. The field energy is given as appropriately 
small value: $E^2_1 / B^2_0 = 10^{-10}$. 
For the moment, $kc / \omega_{pe}$ is fixed to 
unity just for simplicity. 
The field energy exponentially grows in the beginning 
of the run, although other variables have not been much 
changed in a visible manner 
until the field energy reaches at a certain 
level. Then $v_{i0}$ gradually decreases, while 
$K_{e\parallel}$ and $K_i$ increase ($\Omega_i t > 0.2$). 
At the same time, the wave growth slows down and it 
starts damping after $\Omega_i t \sim 0.45$. 
Correspondingly, changing rates of other variables 
decrease and those variables approach some constant 
values when the field energy becomes sufficiently small. 
Such a state is regarded as a saturation 
state and those constant values are defined as 
saturation levels. The saturation state in this system 
is achieved when the field energy is damped out. 
In the analysis we check whether 
the system saturates basically by eye. However, in 
some special cases where the kinetic energies have 
not reached at constant values for long time 
the saturation levels are defined as the values at 
$t=t_0 + 2\Omega^{-1}_i$ where $t_0$ is the time at 
which $v_{i0}$ decreases by $1\%$ from its initial value. 
Since $2\Omega^{-1}_i$ is thought 
to be a typical time scale of shock reformation, 
considering the system evolution longer than this 
time scale may not make sense. 

In Fig.\ref{fig01} $K_{e\perp}$ keeps 
almost constant incontrast to $K_{e\parallel}$ 
or $K_i$. This reflects which kinetic effects 
dominantly work. The linear growth rate, $\gamma_k$, 
can be written by linear superposition of the 
terms such as 
\begin{equation}
 - \xi_{j0} \exp (- \xi^2_{jn}),
\end{equation}
where $\xi_{en} = (\omega_r + n \Omega_e) / 
k_{\parallel} v_{te\parallel}$ and $\xi_{i0} 
= (\omega_r - k v_{i0}) / k v_{ti}$, 
respectively. 
When $|\xi_{jn}| \sim 1$, contribution from 
the corresponding term to the growth rate becomes 
non-negligible. In this sense $\xi_{e0}, \xi_{e1}$, 
and $\xi_{i0}$ can be indicators how strong 
the effects of Landau and cyclotron dampings of 
electrons, and inverse ion Landau damping are, respectively. 
The bottom panel of Fig.\ref{fig01} represents 
values of these indicators. It is now clear that the 
electron Landau damping is the strongest kinetic 
effect and the electron cyclotron damping hardly 
works. The inverse ion Landau damping can also work, 
but may not be so strong as the electron Landau damping 
for this particular case. This can explain the fact 
that increase of $K_{e\parallel}$ is most remarkable, 
while that of $K_{e\perp}$ is unrecognized. 
Contributions from other electron cyclotron harmonics 
($\xi_{en}$ with $n=-1, \pm 2, \pm 3, \pm 4$) are 
further ineffective, and this is confirmed to be true 
for wide range of parameters 
(not shown). Therefore, effects of electron 
cyclotron interactions will be neglected in the 
remaining of the analysis. The assumption that 
$kc / \omega_{pe} = 1$ is fixed is removed hereafter. 

\begin{figure}
\includegraphics[width=8cm]{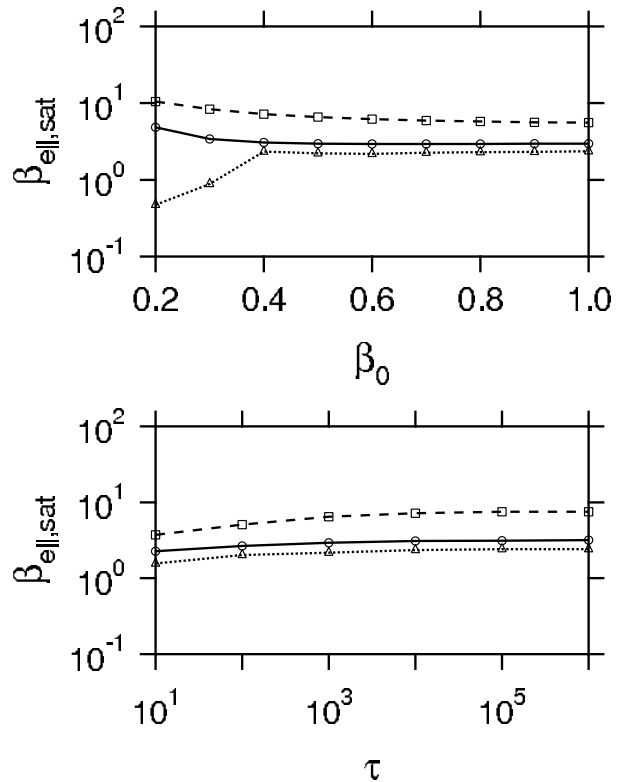}
\caption{Saturation level of normalized parallel electron 
temperatures, $\beta_{e\parallel}$, 
as a function of (upper panel) initial $\beta_0$ 
and (lower panel) $\tau$. The three lines correspond to 
cases with different initial temperature ratio. The solid 
line with circles denotes the case with $\beta_{0e\parallel} 
/ \beta_{0i} = 1$, the dashed line with squares 
$\beta_{0e\parallel} / \beta_{0i} = 3$, and the dotted 
line with triangles $\beta_{0e\parallel} / \beta_{0i} = 1/3$, 
respectively.
\label{fig02}}
\end{figure}
The upper panel of Fig.\ref{fig02} denotes 
saturation levels of normalized electron temperature, 
$\beta_{e\parallel,sat} = 8 \pi n_e T_{e\parallel,sat} 
/ B^2_0$ as a function of the initial 
$\beta_0 = \beta_{0e}+\beta_{0i}$ when $\tau=10^4$. 
In calculating 
the right hand sides of eqs.(\ref{eneea})-(\ref{enei}), 
we chose interval of integration with respect to $k$ 
as $0 \leq kc/\omega_{pe} \leq 3$ and grid size in 
$k$ as $\Delta kc/\omega_{pe} = 0.06$. $\theta_{Bk} 
= 85.5^{\circ}$ is again fixed here. 
In both upper and lower panels the solid line with 
circles, the dashed line with squares, and the 
dotted line with triangles correspond to the cases for 
$\beta_{0e\parallel} / \beta_{0i} = 1, 3$, and $1/3$, 
respectively. $\beta_{e\parallel,sat}$ is essentially 
independent from $\beta_0$ for $\beta_0 > 0.4$ 
indicating that heating process is similar for 
relatively high $\beta_0$. In the low $\beta_0$, on the 
other hand, there is a trend that electron heating is 
suppressed for $\beta_{0e\parallel} / \beta_{0i} = 1/3$. 
For sufficiently low $\beta_{0e\parallel}$, number of resonant 
particles is very little so that it takes extremely long 
time to heat electrons. Then after the very long time, 
heating is suddenly triggered when field energy 
becomes large enough. Such an unnatural time evolution 
occurs because of the assumptions that only the resonant 
wave-particle interactions are taken into account and 
that the 
distribution functions are always Maxellian. Indeed, 
it is known from PIC simulations that effects of 
particle trapping, which have not been 
included in this analysis, become important for 
low $\beta_0$ cases \cite{mat03,mat06b}. 
The initial electron to ion temperature ratio, 
$\beta_{0e\parallel} / \beta_{0i}$, affects 
the saturation levels. For large (small) $\beta_{0e\parallel} 
/ \beta_{0i} (=3 (1/3))$, $\beta_{e\parallel,sat}$ 
is high (low). Comparing with the case of $\beta_{0e\parallel} / 
\beta_{0i} = 1$, $|\xi_{i0}|$ gets closer to unity because $v_{ti}$ 
gets larger when $\beta_{0e\parallel} / \beta_{0i}$ 
becomes 1/3. Hence, the inverse ion Landau 
damping works more effectively so that electron heating 
becomes less dominant. The $\tau$ dependence of 
$\beta_{e\parallel,sat}$ is shown in the lower 
panel of Fig.\ref{fig02} for 
$\beta_0=0.4$. As expected from the linear 
analysis \cite{mat03}, 
dependence on $\tau$ is rather weak. This implies that 
using small $\tau$ in simulations does not lead to 
unrealisitic results for long time evolution of the MTSI.

\begin{figure}
\includegraphics[width=8cm]{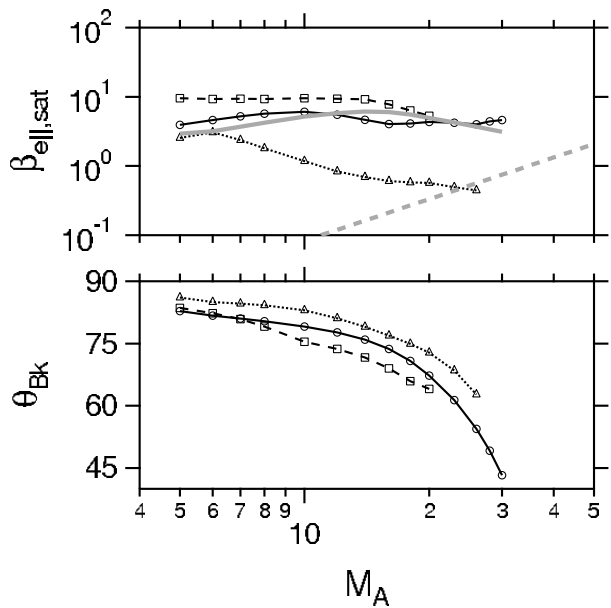}
\caption{Mach number dependence of (upper panel) maximum 
saturation electron temperatures, $\beta_{e\parallel}$, 
and (lower panel) corresponding wave propagation angles, 
$\theta_{Bk}$. The three black lines correspond to cases 
with different initial temperature ratio. Their definition 
is given in the caption of Fig.\ref{fig02}. The 
thick gray solid (dashed) line is obtained from 
eq.(\ref{sat_ql}) (eq.(\ref{sat-ehole})).
\label{fig03}}
\end{figure}
Fig.\ref{fig03} represents $M_A$ dependense 
for $\tau=10^4$ and $\beta_0=0.4$. Three black lines 
(the solid line with circle, the dashed line with squares, 
and the dotted line with triangles) in the upper 
panel denote the maximum saturation 
levels of $\beta_{e\parallel,sat}$ in varying 
$\theta_{Bk}$ again for different initial temperature 
ratios ($\beta_{0e\parallel} / \beta_{0i} = 1, 3, 1/3$). 
The corresponding values of 
$\theta_{Bk}$ is indicated in the bottom panel. 
$\beta_{e\parallel,sat}$ 
does not increase so much, even though the $M_A$ 
gets larger. As already 
mentioned, the system saturates when available field 
energy is lost and no more waves grow. 
Such a state may be achieved when effect of electron 
Landau damping becomes efficient. If it is the case, 
$\eta \xi_{e0} \approx 1$ should be satisfied at the 
saturation state. Here $\eta$ is a factor of the 
order of unity and is empirically determined later. 
By using $v^2_{te\parallel} = T_{e\parallel} / m_e$ and 
$\omega / k_{\parallel} \approx v_{i0} / \cos \theta_{Bk}$, 
\begin{equation}
\label{sat_ql}
 \beta_{e\parallel,sat} \sim {2 \eta^2 M^2_A \over \mu \cos^2 
 \theta_{Bk}} \left( {2 \alpha \over 1 + \alpha} \right)^2.
\end{equation}
The gray thick solid line in Fig.\ref{fig03} is obtained 
by substituting the values of $\theta_{Bk}$ of the solid 
line in the bottom panel into eq.(\ref{sat_ql}). Here, 
$\eta=2.6$ has been assumed. Although 
dependence on initial tempetarure ratio, 
$\beta_{0e\parallel} / \beta_{0i}$, is not trivial 
in eq.(\ref{sat_ql}), this may give a reasonable 
order of magnitude estimate. Remember 
that in the MTSI a cross-field ion beam interacts 
with oblique whistler waves. The whistler waves 
which can interact with the high speed ion beam 
should have appropriately high phase velocities. 
The phase velocity of an oblique whistler wave 
is proportional to $\cos \theta_{Bk}$ \cite{mat03} 
so that a high Mach number shock generates high 
phase velocity, i.e., less oblique, whistler waves. 
This effect appears in the bottom panel. As a result, 
$M_A / \cos \theta_{Bk}$ does not change so much 
even the Mach number varies, so that 
$\beta_{e\parallel,sat}$ remains in the same order 
for wide Mach number regime examined here. Only the 
dotted line with triangles in the upper panel is weakely 
decreasing 
function of the Mach number. In this relatively high 
initial ion temperature case, as mentioned earlier, 
ion kinetic effect seems to be non-negligible. 
As a result, ion heating is more remarkable than in 
the cases with $\beta_{0e\parallel} / \beta_{0i} = 1, 3$. 
Furthermore, because of small growth rate due also to 
the ion kinetic effect, it takes rather long 
time for the system to approach a final saturation 
state. Therefore, the saturation levels were actually 
estimated at $t = t_0 + 2 \Omega^{-1}_i$ as mentioned 
before. 

Note that the condition $\xi_{e0} \approx 1$ 
gives essentially same 
expression given in Ref. 33 
which derives saturation 
energy due to electron trapping. Hence, the expression 
in eq.(\ref{sat_ql}) may be applicable in also the low 
beta case for MTSI. However, because Ref. 33 assumed 
that $\mu \cos^2 \theta_{Bk}$ is constant throughout 
their analyses, their saturation temperature increases 
with $M^2_A$. This might be true if the Mach 
number is sufficiently small and the electrostatic 
approximation is valid. But it is now clear that 
constant $\mu \cos^2 \theta_{Bk}$ assumption is 
invalid at least for $M_A \geq 5$.

\section{Electron Heating Reproduced by 1D PIC Simulation}

\begin{table}
\caption{\label{table1}Injection Plasma Parameters for Simulation}
\begin{tabular}{ccccccc}
\hline
\hline
& $M_{Ain}$ & $\tau$ & $m_i/m_e$ 
& \hspace{0.07cm} $\beta_e$ & \hspace{0.07cm} $\beta_i$ & 
$\theta_{Bn}$ \\
\hline
Run A & 4 & 16 & 625 & \hspace{0.07cm} 0.3 & \hspace{0.07cm} 0.1 & 84 \\
Run B & 6 & 16 & 625 & \hspace{0.07cm} 0.3 & \hspace{0.07cm} 0.1 & 81\\
Run C & 10 & 16 & 625 & \hspace{0.07cm} 0.3 & \hspace{0.07cm} 0.1 & 79 \\
Run D & 8 & 100 & 64 & \hspace{0.07cm} 0.3 & \hspace{0.07cm} 0.1 & 90\\
Run E & 16 & 100 & 64 & \hspace{0.07cm} 0.3 & \hspace{0.07cm} 0.1 & 90\\
Run F & 30 & 100 & 64 & \hspace{0.07cm} 0.3 & \hspace{0.07cm} 0.1 & 90\\
\hline
\hline
\end{tabular}
\end{table}
Here, shock waves are actually reproduced by using one-dimensional 
PIC code and the resultant electron heating through MTSI and BI 
in the transition region is discussed. A shock is produced in 
the simulation domain by using the so-called reflecting wall method. 
An upstream plasma consisting of equal numbers of ions and electrons 
is continuously injected from the left-hand boundary ($x=0$) and 
carries a uniform magnetic field, ${\bf B_0} = (B_{0x}, 0, B_{0z})$, 
and a motional electric field, $E_y = u_{in} B_{0z} / c$, 
where $u_{in}$ is the injection flow velocity in the positive 
$x$-direction. At the right-hand boundary the particles are 
reflected. As a result of the interaction between the 
incoming and reflected particles a shock wave is produced 
and propagates in the negative $x$-direction. Thus the simulation
frame is the downstream rest frame. The grid size is set to be
$\Delta x \approx 0.83 \lambda_{De}$ and 200 particles/cell 
are distributed for both ions and electrons at the injection 
boundary, where $\lambda_{De}$ denotes the electron Debye length. 
Physical parameters are summarized in TABLE \ref{table1}. 
For runs A, B, and C (D, E, and F), MTSI (BI) gets excited 
in the foot during each reformation cycle. 

\begin{figure}
\includegraphics[width=9cm]{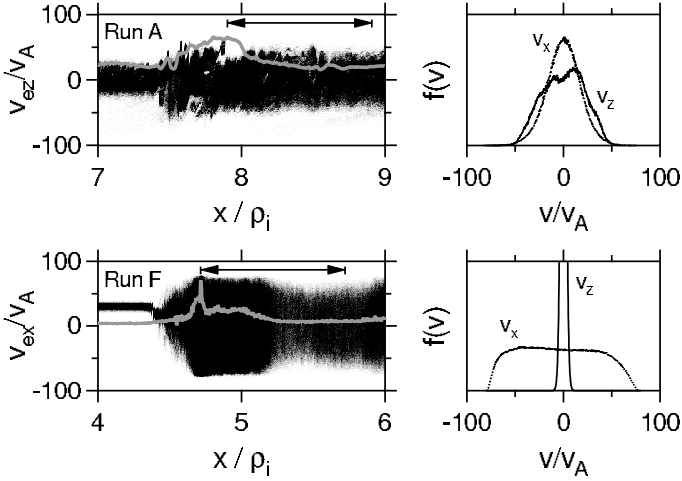}
\caption{Electron heating observed in one dimensional PIC 
simulations for (upper panels) Run A and (lower panels) Run F. 
The left panels show electron phase space distributions in 
(upper left panel) $v_{ez}-x$ and (lower left panel) $v_{ex}-x$. 
The gray solid lines denote profiles of magnetic field $B_z$ 
component. The right panels indicate electron distribution 
functions in $v_{ex}$ and $v_{ez}$ integrated over space 
corresponding to the arrowed regions in the left panels.
\label{fig04}}
\end{figure}
\begin{figure}
\includegraphics[width=8cm]{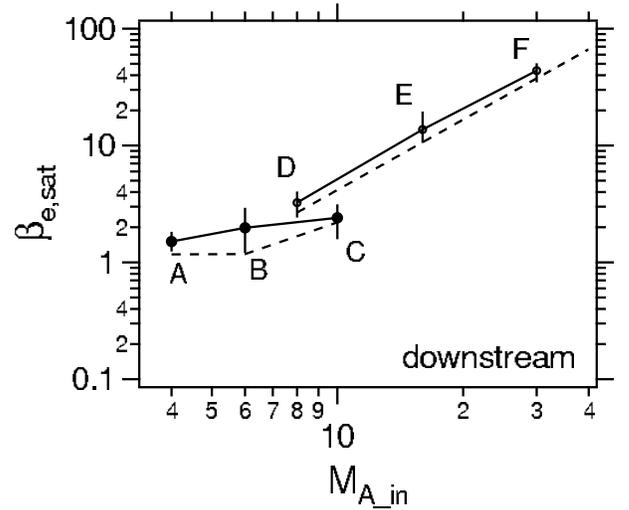}
\caption{Comparison of normalized electron temperatures 
between the PIC simulations and the model analyses. The 
markers labeled A $\sim$ F are estimated as averaged effective 
electron temperatures downstream of the shocks reproduced in 
the PIC simulations. The dashed lines are obtained from the 
model analyses, eq.(\ref{sat_ql}) for A $\sim$ C and 
eq(\ref{sat-ehole}) for D $\sim$ F, respectively.
\label{fig05}}
\end{figure}

Fig.\ref{fig04} shows two examples of snap shots of 
electron phase space $v_z-x$ at $\Omega_i t = 6.02$ for 
Run A (upper left panel) and $v_x-x$ at $\Omega_i t = 8.67$ 
for Run F (lower left panel). The gray solid lines denote 
profiles of magnetic field $B_z$ component. In both runs velocity 
distribution functions corresponding to the downstream areas 
indicated by arrows in the left panels are represented 
in the right panels. Nonadiabatic electron 
heating downstream of the shock is clearly observed. 
Parallel heating is 
superior to the adiabatic perpendicular heating in Run A, 
while strong perpendicular heating, or heating parallel 
to the beam velocity, is seen in Run F. The effective 
temperatures, $K^{eff}$, just downstream of the shocks are 
estimated by using the same definition 
as eqs.(\ref{kpara}) and (\ref{kperp}). We averaged over 
the particles distributed in the region of $x_{os} \leq x 
\leq x_{os}+\rho_i$, where $x_{os}$ is the position of 
the magnetic overshoot and $\rho_i$ the ion cyclotron radius 
defined by upstream flow velocity and ion cyclotron frequency. 
Because of nonstationarity of the shock, $K^{eff}$ is not 
constant in time. Therefore, $K^{eff}$ is also averaged 
in time a few shock reformation periods. The results 
are plotted in Fig.\ref{fig05} as a function of the injection 
Mach number, $M_{Ain}$. 
Here, $K^{eff}$ is normalized to the upstream magnetic 
pressure, $B^2_{01} / 8 \pi$, and 
rewritten as $\beta^{eff}$. For Runs A, B, and C, 
$\beta^{eff}_{e\parallel}$ is almost independent on 
$M_{Ain}$, which is consistent with the extended quasilinear 
analysis shown in the previous section. The dashed line 
just below the points A to C is drawn by using 
eq.(\ref{sat_ql}) with assuming $\alpha=1/3$, $\eta = 1$, 
and $M_A = M_{Ain}$. 

On the other hand, 
$\beta^{eff}_{ex}$ efficiently increases with $M_{Ain}$ for 
Runs D, E, and F, where BI gets excited in the foot. 
Nonlinear development of the BI was well studied in 
the past. Especially, Ref. 10 analyzed it 
in detail by performing one dimensional PIC simulations 
with periodic boundary conditions modeling a part of 
the foot region in a perpendicular shock geometry. They 
showed that when $\tau \gg 1$, the BI produces well 
defined electron holes in its early nonlinear stage and 
some additional heating or acceleration occurs in the 
course of further long time evolution. Although a saturation 
level of the instability is reduced a little in two dimensional 
system, electron trapping still plays an important role 
through oblique modes \cite{ama09b}. 
It is infered 
from these results that size of the electron holes in 
phase space may give a lower limit of $\beta^{eff}_{ex}$. 
The amplitude of an electron hole was estimated by 
assuming that electrostatic wave energy density is 
comparable with drift energy density of electrons in 
the ion's rest frame \cite{ish81,hos02}. 
For our case, it reads 
\begin{equation}
\label{sat-ehole}
 \beta^{eff}_{ex} \sim {8 \over 1 + \alpha} 
 {M^2_A \over \mu^{7/6}},
\end{equation}
where we assume the BI based on electron-reflected ion 
interactions which usually grows faster than the BI 
based on electron-incoming ion interactions so that 
$\Delta u / v_A = 2 M_A / (1+\alpha)$ is used as 
the relative drift velocity. As seen 
in the above expression, $\beta^{eff}_{ex}$ increases 
with $M^2_A$ which is proportional to the system 
free energy. This functional dependence is also derived 
in Refs. 6, 7, and 11. 
One can confirm that eq.(\ref{sat-ehole}) 
may give a good estimate of the lower limit of 
$\beta^{eff}_{ex}$ by plotting the corresponding 
values in Fig.\ref{fig05} for $\alpha = 0.5$ and 
$\mu = 64$ as the dashed line just below the points 
D to F. One should note that 
$M_A$ in eq.(\ref{sat-ehole}) is not necessarily 
coincide with $M_{Ain}$ in Fig.\ref{fig05}. 
Because the $M_A$ indicates the local Mach number 
of the foot in the shock rest frame, while the 
$M_{Ain}$ denotes the injection Mach number in 
the simulation frame, i.e., the downstream rest 
frame. It is also hard to define the local Mach 
number of the foot from the simulation data, 
since the shock is not time stationary. 
Therefore, the dashed line should be 
regarded only as a guide.

\section{Discussions}

In the previous section it was confirmed that the extended 
quasilinear analysis for the MTSI and the trapping analysis 
for the BI give consistent results with the PIC simulations. 
Here, let us compare the results of these two analysis. 
The thick gray dashed line in Fig.\ref{fig03} denotes 
$\beta^{eff}_{ex}$ obtained from eq.(\ref{sat-ehole}) for 
$\mu = 1836, \alpha = 0.5$. This line intersects at 
$M_A \sim 24$ with the dotted line with triangles which is 
$\beta_{e\parallel,sat}$ for the MTSI with 
$\beta_{0e\parallel} / \beta_{0i} = 1/3$. The gray thick 
solid line based on eq.(\ref{sat_ql}) likely to intersect with 
the gray thick dashed line at $M_A \sim 45$. Depending on 
parameters, an actual intersection may occur in the range 
of $20 < M_A < 50$. If 
this is written as $M^*_A$, the dominant electron heating 
process possibly switches from the MTSI for $M_A < M^*_A$ to 
the BI for $M_A > M^*_A$. For relatively low Mach numbers like 
in the earth's bow shock, $\beta_{e\parallel,sat}$ is more 
or less constant with $10^{0 \sim 1}$ because the MTSI 
becomes dominant. If $\beta_{e,sat}=5$ and 
$\tau = 10^4$ are assumed, corresponding electron temperature is 
$\sim 100$eV which is consistent with a typical temperature 
observed downstream of the earth's bow shock. However, 
addiabatic heating due to increases of magnetic field 
and cross shock potential also results in the downstream 
electron temperature similar to this value. This 
may be the reason why remarkable electron heating was 
seldom observed in near earth shocks in the past. But 
one may capture the nonadiabatic parallel electron heating 
if upstream electron beta is low enough ($\beta_{e0} < 0.1$), 
since eq.(\ref{sat_ql}) is still valid for rather low 
beta cases where the trapping effects become essential 
as mentioned in section II. Furthermore, 
as shown in the upper right panel of Fig.\ref{fig04}, 
local temperature anisotropy, $T_{e\parallel}/T_{e\perp} 
> 1$, in a transition region may be observed as a result 
of strong MTSI in some parameter regimes. On the other 
hand, the effective electron temperature significantly 
increases with being proportinal to $M^2_A$ in 
$M_A > M^*_A$. Although this is consistent with the past 
studies \cite{pap88,car88,shi05}, the saturation electron 
temperature estimated in this study seems to be rather 
smaller. This should be because of that the heating process 
in highly nonlinear stage including the second step ion 
acoustic instability is neglected. Electron temperature 
seen in Ref. 10 seems to be about one order higher than 
the estimate given by eq.(\ref{sat-ehole}). This implies 
that an actual $M^*_A$ may appear at a little smaller value.

Consistency of the results of the extended quasilinear 
analysis and the PIC simulations for the MTSI dominant 
cases indicates that the resonant wave-particle interactions 
are essential in electron heating through the MTSI. In 
other wards, nonresonant wave-particle interactions which 
have been neglected in the analysis may not be effective. 
It should also be noted that using unrealistically small 
$\tau$ in PIC simulations is justified as far as the MTSI 
is concerned because of the weak dependence on $\tau$ 
(bottom panel of Fig.2) in contrast to the BI 
dominant systems.

The transition at $M_A = M^*_A$ might not be so drastic 
in a realistic case. 
In the analyses presented here, all other possible 
candidates of microinstabilities have been neglected. 
For example, ECDI (electron cyclotron drift instability) 
is one candidate \cite{mus06} 
and it might become 
important around this critical Mach number, although 
there are some negative indications for the ECDI to 
become dominant. For instance, it is known that 
the saturation level of the ECDI is usually not so 
high \cite{wu84}. 
In addition, if the ECDI gets 
excited simultaneously with the MTSI, the MTSI becomes 
dominant for wide parameter range \cite{mat06b}. 
Furthermore, it is shown that the 
ECDI can be important only when $\tau$ is of the 
order of unity \cite{shi04}. Hence, the Mach number regime where 
the ECDI is dominant seems not to be so wide, but 
we do not remove the possibility that the ECDI 
gives some contributions to electron heating 
around $M^*_A$. Because the ECDI is insensitive 
to the electron Landau damping which may strongly 
affect to the BI in this Mach number regime. 
Further, it is known that the MTSI cannot 
get excited in extremely high Mach number shocks 
like $M_A > 30 \sim 40$ \cite{mat03}. 
There are also other possible 
microinstabilities 
in higher dimensional cases \cite{wu84}. 
Contributions from them should be carefully estimated. 
Some of them have been discussed recently by performing 
two dimensional PIC simulations 
\cite{ohi08,ume08,ama09a}.
The multi-dimansionality may give another important 
contributions to electron heating. Ref. 42 
pointed out that electrons are accelerated in the 
so-called rippled structure which is the ion scale 
structure along the shock surface. Nevertheless, what should 
be emphasized here is that the Mach number dependence 
of the effective electron temperature is systematically 
different between high and low Mach number regimes.

In the present extended quasilinear analysis 
the damping of the wave energy in the late stage (e.g., 
$\Omega_i t > 0.45$ in Fig.\ref{fig01}) may be an artifact. 
Because of the assumption that the distribution function is 
always Maxwellian, the so-called plateau of the 
distribution function is never produced in this system. 
The field energy might saturate earlier at a certain level if 
the plateau is produced. In this regard, it might be 
better that the saturation levels are defined as the 
values at the time when the field 
energy becomes the maximum. However,  the 
resultant saturation levels based on the two definitions, 
i.e., the saturation temperatures estimated at times 
corresponding to the maximum and the later sufficiently 
small field energies, 
are not so much different from each other. Although 
it is possible to solve the quasilinear equation, 
eq.(\ref{qleq}), directly as done in Refs. 39 and 40, 
that is the future work.

The explorations of the inner heliosphere will be underway 
through the BepiColombo mission. The perihelion point of 
the mercury is about 0.3AU where quite high velocity IPSs 
which have not been decerelated are expected to be observed. 
Ref. 43 estimated the propagation speed of one of the IPSs 
observed in August 1972 and concluded that it approached to 
$\sim 2000$ km/s at 0.3AU from the sun. More optimistic 
estimate for the same IPS is given in Ref. 44 as 
$\sim 4000$ km/s at 0.3AU. These results imply that the Mach 
number of this IPS approached to several tens. Therefore, 
the transition of the electron heating efficiency at $M^*_A$, 
if present, will be possibly observed in situ \cite{hos02}.


\begin{acknowledgments}
The author thanks M. Scholer, T. Hada, P. Yoon, and R. Yamazaki for 
useful discussions. 
The PIC simulation was performed by the super computer in ISAS/JAXA 
Sagamihara. This work was supported in part by KAKENHI(19740304).
\end{acknowledgments}

\end{document}